\documentclass[%
 reprint,
 superscriptaddress,
 amsmath,amssymb,
 aps,
pra,
]{revtex4-1}

\usepackage{graphicx}
\usepackage{dcolumn}
\usepackage{bm}

\usepackage{xcolor}
\usepackage{hyperref}
\hypersetup{
	colorlinks=true,
	breaklinks=true,
	linktocpage,
	linkcolor=blue,
	urlcolor=blue,
	citecolor=blue,
	pdfborder={0 0 0}
}
\usepackage{subfigure}
\usepackage{braket}
\usepackage{dsfont}

\newcommand{\expp}[1]{\exp\left\{#1\right\}}

\newcommand{\dx}{\! \! \! \mathrm{d}}
\newcommand{\dmu}{\! \! \mathrm{d}\mu}

\newcommand{\im}{\mathrm{i} \mspace{1mu}} 
\newcommand{\e}{\mathrm{e}} 

\renewcommand{\vector}[1]{\begin{pmatrix}#1\end{pmatrix}} 
\renewcommand{\vec}[1]{\boldsymbol{\mathbf{#1}}}
\newcommand{\hatvec}[1]{\boldsymbol{\mathbf{\hat{#1}}}}

\newcommand{\ma}{m_\mathrm{A}}
\newcommand{\mb}{m_\mathrm{B}}
\newcommand{\mc}{m_\mathrm{C}}
\newcommand{\spina}{{\mathrm{s}_\mathrm{A}}}
\newcommand{\spinb}{{\mathrm{s}_\mathrm{B}}}
\newcommand{\aspina}{{\mathrm{A} \, {\mathrm{s}_\mathrm{A}}}}
\newcommand{\bspinb}{{\mathrm{B} \, {\mathrm{s}_\mathrm{B}}}}

\begin{document}

\title{Relativistic Bell Test within Quantum Reference Frames}

\author{Lucas F. Streiter}
\email{l.streiter@posteo.org} 
\thanks{These two authors contributed equally}
\affiliation{Vienna Center for Quantum Science and Technology (VCQ), Faculty of Physics, University of Vienna, Boltzmanngasse 5, A-1090 Vienna, Austria}
\affiliation{Institute of Quantum Optics and Quantum Information (IQOQI), Austrian Academy of Sciences, Boltzmanngasse 3, A-1090 Vienna, Austria}

\author{Flaminia Giacomini}
\email{fgiacomini@perimeterinstitute.ca}
\thanks{These two authors contributed equally}
\affiliation{Vienna Center for Quantum Science and Technology (VCQ), Faculty of Physics, University of Vienna, Boltzmanngasse 5, A-1090 Vienna, Austria}
\affiliation{Institute of Quantum Optics and Quantum Information (IQOQI), Austrian Academy of Sciences, Boltzmanngasse 3, A-1090 Vienna, Austria}
\affiliation{Perimeter Institute for Theoretical Physics, 31 Caroline St. N, Waterloo, Ontario, N2L 2Y5, Canada}

\author{\v{C}aslav Brukner}
\affiliation{Vienna Center for Quantum Science and Technology (VCQ), Faculty of Physics, University of Vienna, Boltzmanngasse 5, A-1090 Vienna, Austria}
\affiliation{Institute of Quantum Optics and Quantum Information (IQOQI), Austrian Academy of Sciences, Boltzmanngasse 3, A-1090 Vienna, Austria}

\date{\today}

\begin{abstract}
\noindent
A still widely debated question in the field of relativistic quantum information is whether entanglement and the degree of violation of Bell's inequalities for massive relativistic particles are frame independent or not. At the core of this question is the effect that spin gets entangled with the momentum degree of freedom at relativistic velocities. Here, we show that Bell's inequalities for a pair of particles can be maximally violated in a special-relativistic regime, even without any post-selection of the momentum of the particles. To this end, we use the methodology of quantum reference frames, which allows us to transform the problem to the rest frame of a particle, whose state can be in a superposition of relativistic momenta from the viewpoint of the laboratory frame. We show that, when the relative motion of two particles is non-collinear, the optimal measurements for violation of Bell's inequalities in the laboratory frame involve ``coherent Wigner rotations". Moreover, the degree of violation of Bell's inequalities is independent of the choice of the quantum reference frame. Our results open up the possibility of extending entanglement-based quantum communication protocols to relativistic regimes. 
\end{abstract}

\maketitle


\section{Introduction}
Since its formulation in 1964, Bell's theorem has played a crucial role in quantum theory, both for its role in understanding the foundations of the theory, and for its ubiquitous applications in quantum technologies. In 2015, the first experimental \textit{loophole-free} violation of the Clauser-Horne-Shimony-Holt (CHSH) Bell inequality was achieved \cite{Hensen2015,Giustina2015,Shalm2015}, thereby showing that any locally causal description of nature can be ruled out.

So far, Bell's theorem has been verified for massive quantum particles only in the non-relativistic regime \cite{lamehi1976quantum, sakai2006spin, moehring2007entanglement, ritter2012elementary, hofmann2012heralded, rosenfeld2017event, shin2019bell}. In the relativistic regime, entanglement and the violation of Bell's inequalities are still largely discussed in the literature, because of the fact that the spin degrees of freedom lose their coherence due to entanglement with the momentum degree of freedom~\cite{Peres2002}. If coherence were reduced in the special-relativistic regime, protocols involving the violation of Bell's inequalities would break down for high-velocities of the particles, and the special-relativistic corrections in the motion of the particles would result in the introduction of noise. Related to this, there is still ongoing discussion on whether the violation of Bell's inequalities depends on the reference frame: while some authors claim that it is frame-independent \cite{Terashima2002a, Terashima2002,Lee2004,Kim2005,Moradi2009a,Friis2010, castro2012entanglement, castro2015lorentz}, others found it to be frame-dependent \cite{Czachor1997,Terno2003,Ahn2002,Ahn2003,Moon2004,Caban2005,Caban2009,Caban2010,Saldanha2012CHSH, landulfo2009influence, landulfo2010influencephoton, zambianco2019observer}. In fact, Lorentz boosts of the full (spin and momentum) state lead in general to a loss of coherence in the reduced spin states of two particles such that different inertial observers seemingly do not agree on the violation of the CHSH-Bell inequality. Moreover, the theoretical tools to address it in full generality are still to be developed since, with the notable exception of Ref.~\cite{Saldanha2012CHSH}, only quantum particles having a sharp state in momentum have been considered. In Ref.~\cite{Saldanha2012CHSH}, the authors conclude that no maximal violation of Bell's inequalities is possible without postselecting on the momenta of the two particles.

The core of the problem in addressing the question on the violation of Bell's inequalities in the relativistic regime lies in the correct identification of a relativistic spin operator. Several proposals for a relativistic spin operator have been used in the literature to test the violation of Bell's inequalities \cite{Czachor1997,Terashima2002a,Terashima2002,castro2012entanglement, castro2015lorentz, Terno2003,Ahn2002,Ahn2003,Moon2004,Lee2004,Kim2005,Caban2005,Caban2009,Moradi2009a,Caban2010,Friis2010,Saldanha2012CHSH, landulfo2009influence, landulfo2010influencephoton}. 
As a result, it is unknown if one can devise a Bell test for two Dirac particles moving in an arbitrary superposition of relativistic momenta, which would result in a frame-independent statement on the violation of the CHSH-Bell inequality. 

Here, we show that the CHSH-Bell inequality for massive particles in the special-relativistic regime can be maximally violated without postselecting on the momentum of the particles, thereby solving the open problem. Key to this result is the definition of a \emph{quantum reference frame} (QRF) transformation to the rest frame of a particle moving in a quantum superposition of velocities (momenta). Here, by QRF we mean a coordinate system associated to a quantum (physical) system, whose state can be in a superposition or entangled with other surrounding systems. We show that the violation of the CHSH-Bell inequality is independent of the QRF chosen and that, in particular, it can be maximally violated with a specific choice of the initial state and by transforming appropriately the observables from the rest frame to the laboratory frame. Thanks to the operational identification of the observables that maximally violate Bell's inequalities in the relativistic regime, the range of application of the technologies utilising tools from the field of Bell nonlocality, among which Quantum Key Distribution, Quantum Communication Complexity, and device-independent protocols (see Ref.~\cite{brunner2014bell} for a comprehensive review) can be extended to special-relativistic quantum particles. This paves the way for future applications of these techniques in relativistic quantum information.

In Ref.~\cite{Giacomini2019} a formalism was introduced to generalise reference frame transformations to when reference frames are in a quantum relationship with each other, i.e., one QRF is associated to a quantum state from the point of another QRF. In addition, this method enables one to ``jump'' into the rest frame of a system moving in a superposition of velocities. The formalism of Ref.~\cite{Giacomini2019} was generalised in Ref.~\cite{Giacomini2019Spin} to when the particle constituting the reference frame moves in a superposition of relativistic velocities. By introducing a transformation corresponding to the ``superposition of Lorentz boosts'', it was possible to define the spin operationally in the special relativistic regime \cite{Giacomini2019Spin}, via a relativistic Stern-Gerlach experiment. A review of the results obtained when one relativistic particle with spin is considered is given in Appendix~A in \emph{Supplemental Information}. The present work extends the methodology of QRFs to more than one particle to solve the open problem on the violation of Bell's inequalities for a pair of Dirac particles. We note that using the theory of steering and entanglement for qubit states, as well as our representation of the relativistic spin particle as a qubit \cite{Giacomini2019Spin}, one may extend the present schemes to the tasks of steering and entanglement tests for relativistic quantum Dirac particles.

We consider two entangled particles, $\mathrm{A}$ and $\mathrm{B}$, with spin and moving at relativistic velocities. We indicate with $\mathrm{A}$ ($\mathrm{B}$) the external momentum degree of freedom (d.o.f.)~of the particle, and with $\spina$ ($\spinb$) the spin d.o.f.~of the particle. For simplicity, we consider the motion of each particle to be one dimensional, but assume that the two particles, in general, move on a plane, i.e., the velocities need not be collinear. In general, this involves a Wigner rotation under change of QRF.

\section{Collinear motion}

The simplest situation is when the two particles $\mathrm{A}$ and $\mathrm{B}$ move along the same spatial direction, either parallel or anti-parallel, from the point of view of the laboratory frame $\mathrm{C}$. In an arbitrary frame, in which particles of a multi-particle system can have different states of momenta, there is the problem of finding a spin operator which can describe operationally the spin of each particle. Using techniques from QRFs, we build a transformation to the rest frame of one of the two particles, say $\mathrm{A}$. In this frame, the operators describing the spin of $\mathrm{A}$ coincide with the usual Pauli operators, and the spin of particle $\mathrm{B}$ can be described relative to them using a suitable transformation from the rest frame of $\mathrm{B}$ to the one of $\mathrm{A}$ using the results of Ref.~\cite{Giacomini2019Spin}. 

We consider a momentum space representation of the total Hilbert space, where the basis in the rest frame of $\mathrm{A}$ is $\ket{a}_\spina \ket{\pi_B; \Sigma(b)}_\bspinb \ket{\pi_\mathrm{C}}_\mathrm{C}$, with $a$ referring to the spin state of $\mathrm{A}$, $\pi_B$ being the spatial part of the four-momentum of particle $\mathrm{B}$ as seen from $\mathrm{A}$,  $\ket{\pi_B; \Sigma(b)}_\bspinb \equiv \hat{U}_\bspinb(L_{\pi_\mathrm{B}/\mb}) \ket{0; b}_\bspinb$ and $b$ referring to the spin of the Dirac particle $\mathrm{B}$ in its rest frame, as well as the relative momentum $\pi_\mathrm{C}$ between the laboratory $\mathrm{C}$ and the rest frame of $\mathrm{A}$. Notice that $\hat{U}_\bspinb(L_{\pi_\mathrm{B}/\mb})$ is a unitary representation of a pure Lorentz boost acting on particle $\mathrm{B}$ and the Lorentz boost matrix is explicitly given in Appendix~B in \emph{Supplemental Information}.

\begin{figure}[htb]
	\centering
 	\includegraphics[scale=0.2]{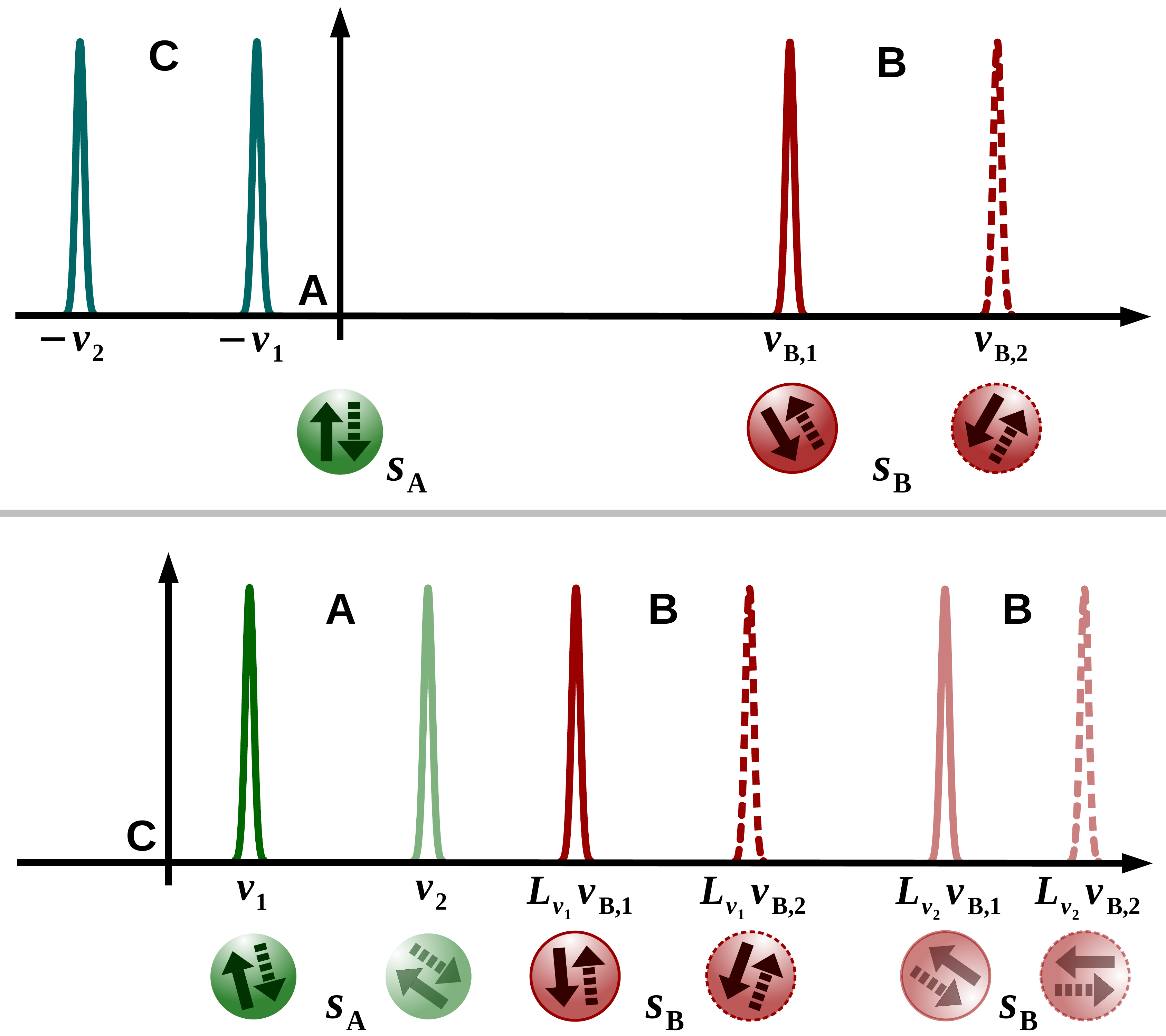}
	\caption{In $\mathrm{A}$'s perspective (above), the spin $\spinb$ of the Dirac particle $\mathrm{B}$ depends on its momentum since $\mathrm{B}$ is moving in a superposition of two sharp relativistic velocities $v_{\mathrm{B},1}$ and $v_{\mathrm{B},2}$. Moreover, the state of the laboratory $\mathrm{C}$ is in a superposition of two relativistic velocities $-v_1$ and $-v_2$ relative to $\mathrm{A}$. In the initial QRF $\mathrm{A}$, the spin $\spina$ and the Dirac particle $\mathrm{B}$ are entangled (similarly to the singlet state $\ket{\Psi^-} = (\ket{ \uparrow  \downarrow} - \ket{ \downarrow \uparrow})/\sqrt{2}$) which is illustrated by the correlation between the dashed and between the solid arrows. The QRF transformation $\hat{S}_2$ from A to C coherently boosts the two Dirac particles by the velocity of $\mathrm{C}$ and outputs the perspective of the laboratory (below). In the laboratory frame $\mathrm{C}$, the two Dirac particles $\mathrm{A}$ and $\mathrm{B}$ are entangled and both spin $\spina$ and $\spinb$ depend on the corresponding momentum d.o.f.~$\mathrm{A}$ and $\mathrm{B}$.} 
	\label{fig: Collinear Motion}
\end{figure}

From the point of view of particle $\mathrm{A}$, the total state is described by
\begin{equation} \label{eq: General State Rest Frame A}
\ket{\psi}^{|\mathrm{A}}_{\spina \bspinb \mathrm{C}} = \ket{\eta}_{\spina \bspinb} \ket{\phi}_\mathrm{C}
\end{equation}
where 
\begin{equation} 
\ket{\eta}_{\spina \bspinb} = \sum_{a,b} c_{ab} \int \dmu_\mathrm{B}(\pi_\mathrm{B}) \ \eta(\pi_\mathrm{B}) \ket{a}_\spina \ket{\pi_\mathrm{B}; \Sigma(b)}_\bspinb \label{eq: initial state rest frame A}
\end{equation}
is an arbitrary state of spin $\spina$ and particle $\mathrm{B}$ and
\begin{equation}
\ket{\phi}_\mathrm{C} = \int \dmu_\mathrm{C}(\pi_\mathrm{C}) \phi(\pi_\mathrm{C}) \ket{\pi_\mathrm{C}}_\mathrm{C} \label{eq: laboratory state}
\end{equation}
is an arbitrary state of  laboratory $\mathrm{C}$. Here, the (1+1)-momentum is $\pi^\mu_\mathrm{k} = (\sqrt{m_\mathrm{k}^2 c^2 + \pi_\mathrm{k}^2}, \pi_\mathrm{k})$ and the Lorentz-invariant integration measure is $ \  \dmu_\mathrm{k}(\pi_\mathrm{k}) = \tfrac{\dx \pi_\mathrm{k} }{4 \pi \sqrt{m_\mathrm{k}^2 c^2 + \pi_\mathrm{k}^2}}$, with $\mathrm{k}= \mathrm{B, C}$. Hence, in $\mathrm{A}$'s perspective the Dirac particle $\mathrm{B}$ moves in a superposition of momenta and it can be entangled with the state of spin $\spina$. This situation is graphically illustrated in Fig.~\ref{fig: Collinear Motion}. 

The QRF transformation $\hat{S}_2: \mathcal{H}^{|\mathrm{A}}_\spina \otimes \mathcal{H}^{|\mathrm{A}}_\bspinb \otimes \mathcal{H}^{|\mathrm{A}}_\mathrm{C} \mapsto \mathcal{H}^{|\mathrm{C}}_\aspina \otimes \mathcal{H}^{|\mathrm{C}}_\bspinb$ to the laboratory frame is a Lorentz boost of particle $\mathrm{B}$, controlled on the velocity of the laboratory from the perspective of A, composed with the QRF transformation $\hat{S}_\mathrm{L}$ introduced in Ref.~\cite{Giacomini2019Spin} and reviewed in Appendix~A in \emph{Supplemental Information}. Specifically, it is given by
\begin{equation} \label{eq: S2}
\hat{S}_2 = \hat{S}_\mathrm{L} \ \hat{U}_\bspinb(L_{-\hat{\pi}_\mathrm{C}/\mc}) ,
\end{equation}
where $\hat{S}_\mathrm{L}$ (defined in Appendix~A in \emph{Supplemental Information}) corresponds to the QRF transformation from the rest frame of $\mathrm{A}$ to the laboratory frame C acting on the spin $\spina$ and the momentum d.o.f.~of the laboratory $\mathrm{C}$. Specifically, $\hat{S}_\mathrm{L}$ acts on an arbitrary element of the basis of the total Hilbert space of the spin $\spina$ and on the momentum of $\mathrm{C}$ as $\hat{S}_\mathrm{L} \ket{a}_\spina \ket{\pi}_\mathrm{C} = \ket{-\tfrac{\ma}{\mc} \pi; \Sigma(a)}_\aspina$. In addition, $\hat{U}_\bspinb(L_{-\hat{\pi}_\mathrm{C}/\mc})$ is a unitary representation of a pure Lorentz boost acting on the Dirac particle $\mathrm{B}$, where the boost parameter is promoted to an operator. Since two successive collinear Lorentz boosts are a pure Lorentz boost, the action of the QRF transformation $\hat{S}_2$ on the basis is
\begin{equation} \label{eq: action S2 on basis elements}
\begin{split}
\hat{S}_2 & \ket{a}_\spina \ket{\pi_B; \Sigma(b)}_\bspinb \ket{\pi_\mathrm{C}}_\mathrm{C} \\
& = \ket{-\tfrac{\ma}{\mc} \pi_\mathrm{C}; \Sigma(a)}_\aspina \ket{L\pi_\mathrm{B}; \Sigma'(b)}_\bspinb ,
\end{split}
\end{equation}
where in $\mathrm{C}$'s perspective the spin of $\mathrm{A}$ is entangled with its momentum d.o.f.~and $\mathrm{B}$ propagates with the boosted momentum $L\pi_\mathrm{B}$. $L\pi_\mathrm{B}$ refers to the spatial part of $\left(L_{-\pi_\mathrm{C}/\mc}\right)^{\mu}_{\ \nu} \pi_\mathrm{B}^\nu \equiv p_\mathrm{B}^\mu$ where $\pi^\nu_\mathrm{B} = (\sqrt{\mb^2 c^2 + \pi_\mathrm{B}^2}, \pi_\mathrm{B})$. Notice that the transformation acts nontrivially on the spin state $\spinb$, and that, as a result, $\Sigma'$ depends on the momentum. Furthermore, after the transformation to the laboratory frame, the relative state of the laboratory from the perspective of $\mathrm{A}$ is transformed to the relative state of $\mathrm{A}$ from the perspective of the laboratory $\mathrm{C}$, in line with the formalism for QRFs \cite{Giacomini2019, Giacomini2019Spin} (see also Appendix~A in \emph{Supplemental Information}).

The state of $\mathrm{A}$ and $\mathrm{B}$ in the laboratory frame is found by transforming the state in Eq.~\eqref{eq: General State Rest Frame A} with the transformation in Eq.~\eqref{eq: S2}, i.e., $\ket{\psi}^{|\mathrm{C}}_{\aspina \bspinb} = \hat{S}_2 \ket{\psi}^{|\mathrm{A}}_{\spina \bspinb \mathrm{C}}$, where
\begin{equation} \label{eq: LAB State Collinear}
\begin{split}
\ket{\psi}^{|\mathrm{C}}_{\aspina \bspinb} = & \sum_{a,b} c_{ab} \int \dmu_\mathrm{A}(p_\mathrm{A}) \ \ \! \dmu_\mathrm{B}(p_\mathrm{B}) \ \eta(L^{-1} p_\mathrm{B}) \\ 
& \phi\left(- \tfrac{\mc}{\ma} p_\mathrm{A}\right) \ket{p_\mathrm{A};\Sigma(a)}_\aspina \ket{p_\mathrm{B}; \Sigma'(b)}_\bspinb ,
\end{split}
\end{equation}
$p_\mathrm{A} \equiv - \tfrac{\ma}{\mc} \pi_\mathrm{C}$, $p_\mathrm{B}$ refers to the spatial part of $p_\mathrm{B}^\mu \equiv (L_{-\pi_\mathrm{C}/m_\mathrm{C}})^\mu_{\ \nu} \pi_\mathrm{B}^\nu$ and $L^{-1}p_\mathrm{B}$ refers to the spatial part of $\pi_\mathrm{B}^\mu \equiv (L_{p_\mathrm{A}/m_\mathrm{A}}^{-1})^\mu_{\ \nu} p_\mathrm{B}^\nu$, and $\ \dmu_\mathrm{k}(p_\mathrm{k})$, $\mathrm{k}= \mathrm{A, B}$ is the Lorentz-invariant integration measure previously introduced. Thus, in $\mathrm{C}$'s perspective the two Dirac particles $\mathrm{A}$ and $\mathrm{B}$ are entangled and their spin d.o.f., $\spina$ and $\spinb$, are momentum-dependent, see Fig.~\ref{fig: Collinear Motion}.

In the rest frame of the Dirac particle $\mathrm{A}$ a proper spin observable for its spin state $\spina$ is given (as in non-relativistic quantum mechanics) by the Pauli operators $\hat{\sigma}^i_\spina$. In analogy to the one particle case \cite{Giacomini2019Spin}, the observable $\hat{\Xi}^j_{\hat{\pi}_\mathrm{B}} = \hat{S}_L (\mathds{1}_\mathrm{B} \otimes \hat{\sigma}^j_{\spinb})\hat{S}_L^\dagger$, where the $\hat{S}_L$ operator acts on $\mathrm{B}$, is used as relativistic spin observable for $\mathrm{B}$ in the rest frame of A. Consequently, the joint spin measurement in $\mathrm{A}$'s rest frame is described by
\begin{equation} \label{eq: Bell Observable Perspective A}
 \vec{x} \cdot \hatvec{\sigma}_{\spina} \otimes \vec{y} \cdot \hatvec{\Xi}_{\hat{\pi}_\mathrm{B}} \otimes \mathds{1}_\mathrm{C} = \sum_{i,j} x^i y^j  \ \hat{\sigma}_{\spina}^i \otimes \hat{\Xi}_{\hat{\pi}_\mathrm{B}}^j \otimes \mathds{1}_\mathrm{C} ,
\end{equation}
where the vectors $\vec{x}$ and $\vec{y}$ denote the measurement settings. Evaluating Eq.~\eqref{eq: Bell Observable Perspective A} for the general state of Eq.~\eqref{eq: General State Rest Frame A} requires the calculation of the corresponding entries of the correlation tensor, which is detailed in Appendix~C in \emph{Supplemental Information}. For the CHSH-Bell inequality and the general definition, see Appendix~D in \emph{Supplemental Information}. With this result, it is easy to show that the CHSH-Bell inequality is maximally violated for the state 
\begin{equation} \label{eq: Signlet State in perspective A}
	\ket{\psi^-}^{|\mathrm{A}}_{\spina \bspinb \mathrm{C}} = \ket{\eta^-}_{\spina \bspinb} \ket{\phi}_\mathrm{C}
\end{equation}
where \\ $\ket{\eta^{-}}_{\spina \bspinb} = \sum_{\lambda=\pm z} c_\lambda \int \dmu_\mathrm{B}(\pi_\mathrm{B}) \ \eta(\pi_\mathrm{B}) \ket{\lambda}_\spina \ket{\pi_\mathrm{B}; \Sigma(-\lambda)}_\bspinb$ and $c_{\pm z} = \pm 1 / \sqrt{2}$.

The Bell observable in the laboratory frame is (see Appendix~E in \emph{Supplemental Information} for details) 
\begin{equation} \label{eq: Bell Observable Collinear}
\hat{S}_2 \left( \vec{x} \cdot \hatvec{\sigma}_{\spina} \otimes \vec{y} \cdot \hatvec{\Xi}_{\hat{\pi}_\mathrm{B}} \otimes \mathds{1}_\mathrm{C} \right) \hat{S}_2^\dagger = \vec{x} \cdot \hatvec{\Xi}_{\hat{p}_\mathrm{A}} \otimes \vec{y} \cdot \hatvec{\Xi}_{\hat{p}_\mathrm{B}} . 
\end{equation}
Note that the observables on $\mathrm{A}$ and $\mathrm{B}$ are factorised, and that the two spin measurements on the two Dirac particles $\mathrm{A}$ and $\mathrm{B}$ can be performed independently in space-like separated regions; thus, we have shown that it is possible to extend the Bell test to the special-relativistic domain in the laboratory $\mathrm{C}$. Due to the unitarity of the QRF transformation $\hat{S}_2$, the amount of the violation of the CHSH-Bell inequality is QRF-independent. In addition, we find that, in this configuration, the measurement settings $\vec{x}$ as well as $\vec{y}$ do not change after the QRF transformation (albeit the observables measured are changed and involve spin and momenta degrees of freedom on both sides).

\section{Non-collinear relative motion}
The previous treatment can be generalised to scenarios where the the Dirac particle $\mathrm{B}$ and the laboratory $\mathrm{C}$ move along different spatial directions from the QRF of particle $\mathrm{A}$. For simplicity, we demand the motion of each particle to be one dimensional, so that the angle $\xi$ between the momentum of the Dirac particle $\vec{\pi}_\mathrm{B}$ and the laboratory $\vec{\pi}_\mathrm{C}$ is fixed, i.e., the two systems do not move in a superposition of directions. Without loss of generality, we replace the previously considered 1-dimensional momenta by 3-dimensional vectors according to $\pi_\mathrm{B} \to \vec{\pi}_\mathrm{B} = (\pi_\mathrm{B}, 0, 0) = \pi_\mathrm{B} \vec{e}_x$
and $\pi_\mathrm{C} \to \vec{\pi}_\mathrm{C} = \pi_\mathrm{C} \vec{u}$, where $\vec{u} = (u_x,u_y,0)$ and $|\vec{u}| = 1$. Since $\xi$ is fixed, we can treat the motion of each particle as one-dimensional by introducing the projections $\pi_\mathrm{B} \equiv \vec{\pi}_\mathrm{B} \cdot \vec{e}_x$ and $\pi_\mathrm{C} \equiv \vec{\pi}_\mathrm{C} \cdot \vec{u}$. Thanks to this condition, the state in the QRF of particle $\mathrm{A}$ is formally written in the same way as in the previous case, i.e., $\ket{\vec{\psi}}^{|\mathrm{A}}_{\spina \bspinb \mathrm{C}} = \ket{\vec{\eta}}_{\spina \bspinb} \ket{\vec{\phi}}_\mathrm{C}$, where state vectors are now printed in bold to differentiate between the collinear case. Crucially, this allows us to conclude immediately that the CHSH-Bell inequality is violated in the rest frame of particle $\mathrm{A}$ if we consider the quantum state in Eq.~\eqref{eq: Signlet State in perspective A} and the observable $\vec{x} \cdot \hatvec{\sigma}_{\spina} \otimes \vec{y} \cdot \hatvec{\Xi}_{\hatvec{\pi}_\mathrm{B}} \otimes \mathds{1}_\mathrm{C}$
with proper measurement settings $\vec{x}$ and $\vec{y}$.

However, the description in the laboratory frame is different now due to an additional Wigner rotation \cite{Wigner1989, Weinberg1995} of the spin d.o.f.~$\spinb$. The Wigner rotation appears, in contrast to the previous discussion, because the laboratory $\mathrm{C}$ and the Dirac particle $\mathrm{B}$ do not move along the same spatial direction from the viewpoint of the rest frame of $\mathrm{A}$. Specifically, with the (1+3)-dimensional extension of the previous QRF transformation $\hat{S}_2 \equiv \hat{S}_\mathrm{L} \ \hat{U}_\bspinb(L_{-\hatvec{\pi}_\mathrm{C}/\mc})$, we get
\begin{equation}
\begin{split}
& \hat{S}_2 \ket{a}_\spina \ket{\vec{\pi}_\mathrm{B}; \Sigma(b)}_\bspinb \ket{\vec{\pi}_\mathrm{C}}_\mathrm{C} = \ket{-\tfrac{\ma}{\mc} \vec{\pi}_\mathrm{C}; \Sigma(a)}_\aspina  \\
& \quad \quad \quad \ \ \hat{U}_\bspinb(L_{-\vec{\pi}_\mathrm{C}/\mc}) \hat{U}_\bspinb(L_{\vec{\pi}_\mathrm{B}/\mb}) \ket{0; b}_\bspinb .
\end{split}
\end{equation}
The two successive non-collinear ($\vec{\pi}_\mathrm{B} \nparallel \vec{\pi}_\mathrm{C}$) Lorentz boosts on particle $\mathrm{B}$ result in a Wigner rotation of its spin and a pure boost taking its momentum to $\vec{p}_\mathrm{B}$, where $\left(L_{-\vec{\pi}_\mathrm{C}/\mc}\right)^{\mu}_{\ \nu} \pi_\mathrm{B}^\nu \equiv p_\mathrm{B}^\mu \equiv (p_\mathrm{B}^0,\vec{p}_\mathrm{B})$, according to
\begin{equation}
\begin{split}
\hat{U}_\bspinb&(L_{-\vec{\pi}_\mathrm{C}/\mc})  \hat{U}_\bspinb(L_{\vec{\pi}_\mathrm{B}/\mb}) \ket{0; b}_\bspinb \\
& = \hat{U}_\bspinb(L_{\vec{p}_\mathrm{B}/\mb}) \left[\mathds{1}_\mathrm{B} \otimes \hat{R}_\spinb(\vec{\Omega})\right] \ket{0; b}_\bspinb \\
& = \ket{\vec{p}_\mathrm{B}; \Sigma(R_{\vec{\Omega}}(b))}_\bspinb,
\end{split}
\end{equation}
where $\hat{R}_\spinb(\vec{\Omega})$ is the Wigner rotation about the axis orthogonal to the directions of the two boosts $\hat{U}_\bspinb(L_{-\vec{\pi}_\mathrm{C}/\mc})$ and $\hat{U}_\bspinb(L_{\vec{\pi}_\mathrm{B}/\mb})$. In general, the spin is rotated in a superposition of angles, depending on the eigenvalue of the operators $\hatvec{\pi}_\mathrm{C}$ and $\hatvec{\pi}_\mathrm{B}$ when they act on a basis of the total Hilbert space.

The rotation is specified through $\vec{\Omega}\equiv \Omega \vec{n}, |\vec{n}|=1$. More specifically, the spin $\spinb$ is rotated about the axis
\begin{equation}
\vec{n} = \dfrac{\vec{\pi}_\mathrm{B} \times \vec{\pi}_\mathrm{C}}{|\vec{\pi}_\mathrm{B} \times \vec{\pi}_\mathrm{C}|} = \vec{e}_z \equiv (0, 0, 1)
\end{equation}
by the angle $\Omega$ which is given by
\begin{equation} \label{eq:OmegainA}
\cos \Omega = \dfrac{1 + \gamma_{\vec{\pi}_\mathrm{B}} + \gamma_{\vec{\pi}_\mathrm{C}} + \gamma_{L\vec{\pi}_\mathrm{B}}}{(1+\gamma_{\vec{\pi}_\mathrm{B}})(1+\gamma_{\vec{\pi}_\mathrm{C}})(1+\gamma_{L\vec{\pi}_\mathrm{B}})} - 1 ,
\end{equation}
where $L\vec{\pi}_\mathrm{B}$ refers to the spatial part of $\left(L_{-\vec{\pi}_\mathrm{C}/\mc}\right)^{\mu}_{\ \nu} \pi_\mathrm{B}^\nu$, $\gamma_{\vec{\pi}_\mathrm{K}} = \sqrt{1+\tfrac{\vec{\pi}_\mathrm{K}^2}{m_\mathrm{K}^2 c^2}}$, $\mathrm{K}= \mathrm{B}, \mathrm{C}$,
$\gamma_{L\vec{\pi}_\mathrm{B}} = \gamma_{\vec{\pi}_\mathrm{B}} \gamma_{\vec{\pi}_\mathrm{C}} (1-\vec{\beta}_{\vec{\pi}_\mathrm{B}} \cdot \vec{\beta}_{\vec{\pi}_\mathrm{C}})$ and $\vec{\beta}_{\vec{\pi}_\mathrm{k}} \equiv \tfrac{\vec{\pi}_\mathrm{k}}{\sqrt{m_\mathrm{k}^2 c^2 + \vec{\pi}_\mathrm{k}^2}}$.

Consequently, the rotation operator is $\hat{R}(\vec{\Omega}) = \e^{- \im \vec{\Omega} \cdot \hatvec{\sigma}/2}= \mathds{1} \cos(\Omega/2) - \im \hat{\sigma}_z \sin(\Omega/2)$. Thus, the axis of rotation is fixed and the rotation angle $\Omega$ depends on the relative orientation as well as on the magnitude of the two momenta $\vec{\pi}_\mathrm{B}$ and $\vec{\pi}_\mathrm{C}$, i.e.~$\Omega = \Omega(\vec{\pi}_\mathrm{B},\vec{\pi}_\mathrm{C})$. This implies that the Wigner rotation is controlled on the momentum of the two particles. 

Starting in $\mathrm{A}$'s rest frame with the state
\begin{equation}
\ket{\vec{\psi}}^{|\mathrm{A}}_{\spina \bspinb \mathrm{C}} = \ket{\vec{\eta}}_{\spina \bspinb} \ket{\vec{\phi}}_\mathrm{C}
\end{equation}
where $\ket{\vec{\eta}}_{\spina \bspinb} = \sum_{a,b} c_{ab} \int \dmu_\mathrm{B}(\pi_\mathrm{B}) \ \eta(\vec{\pi}_\mathrm{B}) \ket{a}_\spina \ket{\vec{\pi}_\mathrm{B}; \Sigma(b)}_\bspinb $ and $\ket{\vec{\phi}}_\mathrm{C} = \int \dmu_\mathrm{C}(\pi_\mathrm{C}) \phi(\vec{\pi}_\mathrm{C}) \ket{\vec{\pi}_\mathrm{C}}_\mathrm{C}$ we obtain the total state in the laboratory frame
\begin{equation} \label{eq: LAB state NonCollinear}
\begin{split}
& \ket{\vec{\psi}}^{|\mathrm{C}}_{\aspina \bspinb} = \hat{S}_2 \ket{\vec{\psi}}^{|\mathrm{A}}_{\spina \bspinb \mathrm{C}} =\sum_{a,b} c_{ab} \int \dmu_\mathrm{A}(p_\mathrm{A}) \ \ \! \dmu_\mathrm{B}(p_\mathrm{B}) \  \\
& \eta(L^{-1} \vec{p}_\mathrm{B})  \phi\left(- \tfrac{\mc}{\ma} \vec{p}_\mathrm{A}\right)  \ket{\vec{p}_\mathrm{A};\Sigma(a)}_\aspina \ket{\vec{p}_\mathrm{B}; \Sigma(R_{\vec{\Omega}}(b))}_\bspinb
\end{split}
\end{equation}
where $\vec{p}_\mathrm{A} \equiv - \tfrac{\ma}{\mc} \vec{\pi}_\mathrm{C}$ and $L^{-1}\vec{p}_\mathrm{B}$ refers to the spatial part of $\pi_\mathrm{B}^\mu \equiv (L_{\vec{p}_\mathrm{A}/m_\mathrm{A}}^{-1})^\mu_{\ \nu} p_\mathrm{B}^\nu$. With this, we find $\vec{\Omega} = \vec{\Omega}(\vec{p}_\mathrm{A},\vec{p}_\mathrm{B})$ in the laboratory frame $\mathrm{C}$, where the rotation is also about the $z$-axis and 
\begin{equation} \label{eq: Wigner Angle Omega p A B}
\cos \Omega = \dfrac{1 + \gamma_{\vec{p}_\mathrm{A}} + \gamma_{\vec{p}_\mathrm{B}} + \gamma_{L\vec{p}_\mathrm{B}}}{(1+\gamma_{\vec{p}_\mathrm{A}})(1+\gamma_{\vec{p}_\mathrm{B}})(1+\gamma_{L\vec{p}_\mathrm{B}})} - 1,
\end{equation}
where the notation is consistent with the one adopted in Eq.~\eqref{eq:OmegainA}.

The calculation of the Bell observables in the laboratory frame $\hat{G}^{\vec{x} \vec{y}}_{|\mathrm{C}} \equiv \hat{S}_2 (\vec{x} \cdot \hatvec{\sigma}_{\spina} \otimes \vec{y} \cdot \hatvec{\Xi}_{\hatvec{\pi}_\mathrm{B}} \otimes \mathds{1}_\mathrm{C}) \hat{S}_2^\dagger$, carried out in detail in Appendix~F in \emph{Supplemental Information}, yields
\begin{equation} \label{eq: Wigner Rot Observable}
\begin{split}
\hat{G}^{\vec{x} \vec{y}}_{|\mathrm{C}} =&  \left( \vec{x} \cdot \hatvec{\Xi}_{\hatvec{p}_\mathrm{A}} \otimes \mathds{1}_\bspinb \right)  \\
& \sum_j \left(y_j^R(\hatvec{p}_\mathrm{A},\hatvec{p}_\mathrm{B}) \otimes \mathds{1}_\spina \otimes \mathds{1}_\spinb \right) \left(\mathds{1}_\aspina \otimes \hat{\Xi}^j_{\hatvec{p}_\mathrm{B}} \right)  
\end{split}
\end{equation}
where $y_j^R(\hatvec{p}_\mathrm{A},\hatvec{p}_\mathrm{B})$ refers to the $j$-th component of $\vec{y}^R(\hatvec{p}_\mathrm{A},\hatvec{p}_\mathrm{B}) = \vec{y} \cos[\Omega(\hatvec{p}_\mathrm{A},\hatvec{p}_\mathrm{B})] + \vec{n} (\vec{n} \cdot \vec{y}) (1-\cos[\Omega(\hatvec{p}_\mathrm{A},\hatvec{p}_\mathrm{B})]) + (\vec{n} \times \vec{y}) \sin[\Omega(\hatvec{p}_\mathrm{A},\hatvec{p}_\mathrm{B})]$. Notice that in the case of sharp momenta $\vec{p}_\mathrm{A}$ and $\vec{p}_\mathrm{B}$ the rotation is specified by a single angle, however, if the particles move in a superposition of momenta, the measurement setting $\vec{y}^R$ is "coherently rotated" with respect to its initial setting $\vec{y}$. In order to measure the observable of Eq.~\eqref{eq: Wigner Rot Observable}, one observer (Alice) measures a local observable on particle $\mathrm{A}$, but another (Bob) measures the spin $\spinb$ along a direction depending on $\mathrm{A}$'s momentum, because $\hat{G}^{\vec{x} \vec{y}}_{|\mathrm{C}} \neq \hat{G}^{\vec{x}}_{\aspina} \otimes \hat{G}^{\vec{y}^R}_{\bspinb}$. Instead, the measurement setting $\vec{y}^R(\hatvec{p}_\mathrm{A},\hatvec{p}_\mathrm{B})$ depends on both momenta of the particles $\mathrm{A}$ and $\mathrm{B}$. Nonetheless, the fact that the operators no longer factorise does not violate the locality assumption of Bell's theorem: it is possible to devise a Bell experiment in which the events, consisting of choosing the settings for the spin and observing spin outcome in the two laboratories, are spacelike separated. Specifically, Alice could entangle an auxiliary system $\mathrm{M}$ with the momentum of her particle before choosing the settings, and then send $\mathrm{M}$ to Bob. In that case, the observable becomes
\begin{equation} 
\begin{split}
\hspace{-.3em}
\hat{G}^{\vec{x} \vec{y}}_{|\mathrm{C}} =& \left[ \vec{x} \cdot \hatvec{\Xi}_{\hatvec{p}_\mathrm{A}} \otimes \mathds{1}_\mathrm{\bspinb} \right] \left[\mathds{1}_\mathrm{\aspina} \otimes \vec{y}^R(\hatvec{p}_\mathrm{M},\hatvec{p}_\mathrm{B})\cdot \hatvec{\Xi}_{\hatvec{p}_\mathrm{B}} \right] \\
=& \ \hat{G}^{\vec{x}}_{\aspina} \otimes \hat{G}^{\vec{y}^R}_{\bspinb \mathrm{M}} \ ,
\end{split}
\end{equation}
where $\vec{x} \cdot \hatvec{\Xi}_{\hatvec{p}_\mathrm{A}}$ and $\vec{y}^R(\hatvec{p}_\mathrm{M},\hatvec{p}_\mathrm{B})\cdot \hatvec{\Xi}_{\hatvec{p}_\mathrm{B}}$ are the scalar products between the measurement settings in the reference frame of $\mathrm{C}$ and the spin operators of the two particles. In the rest frame of $\mathrm{A}$, we consider the measurement settings $\vec{x}_1 = (0,1,1)/\sqrt{2}$, $\vec{x}_2 = (0,1,-1)/\sqrt{2}$, $\vec{y}_1 = (0,1,0)$ and $\vec{y}_2 = (0,0,1)$. Hence, the rotated measurement settings in the laboratory frame $\vec{y}^R(\hatvec{p}_\mathrm{M},\hatvec{p}_\mathrm{B})$ are 
\begin{equation}
\vec{y}_1^R = \vector{-\sin \left[\Omega(\hatvec{p}_\mathrm{M},\hatvec{p}_\mathrm{B})\right] \\ \cos \left[\Omega(\hatvec{p}_\mathrm{M},\hatvec{p}_\mathrm{B})\right] \\ 0}  \quad \textmd{and} \quad \vec{y}_2^R = \vector{0 \\ 0 \\ 1} \equiv \vec{y}_2.
\end{equation}
This rotated setting is obtained simply by inserting $\vec{y}_1$ and $\vec{y}_2$ in $\vec{y}^R(\hatvec{p}_\mathrm{M},\hatvec{p}_\mathrm{B}) = \vec{y} \cos[\Omega(\hatvec{p}_\mathrm{M},\hatvec{p}_\mathrm{B})] + \vec{n} (\vec{n} \cdot \vec{y}) (1-\cos[\Omega(\hatvec{p}_\mathrm{M},\hatvec{p}_\mathrm{B})]) + (\vec{n} \times \vec{y}) \sin[\Omega(\hatvec{p}_\mathrm{M},\hatvec{p}_\mathrm{B})]$ where $\vec{n} = \vec{e}_z$.

\section{Conclusions}
We have shown how to devise a relativistic Bell test with Dirac particles, by introducing operationally well-defined spin operators. Key to the result is the introduction of a transformation to ``jump'' to the rest frame of a general quantum system, in the case where two Dirac particles are moving in a superposition of relativistic velocities. This transformation is built by making use of a formalism to describe physics from the perspective of a quantum reference frame introduced in Refs.~\cite{Giacomini2019,Giacomini2019Spin}. We hence settle the controversy on the violation of Bell's inequalities in different (quantum) reference frames, and by providing an operational identification of the observables that maximally violate Bell's inequalities in a special-relativistic setting, we show that, regardless of what the state of the external degrees of freedom of the particle is, such a maximal violation can be always achieved for particles moving in a superposition of relativistic velocities. This paves the way for the extension of known quantum information technologies, such as Quantum Communication Complexity, Quantum Key Distribution, and device-independent protocols, to massive relativistic spin-1/2 particles moving in a superposition of velocities.

\begin{acknowledgments}
F.G. acknowledges support from Perimeter Institute for Theoretical Physics. Research at Perimeter Institute is supported in part by the Government of Canada through the Department of Innovation, Science and Industry Canada and by the Province of Ontario through the Ministry of Colleges and Universities. \v{C}. B. acknowledges the support from the research platform TURIS, from the European Commission via Testing the Large-Scale Limit of Quantum Mechanics  (TEQ)  (No.  766900)  project, and from the Austrian-Serbian bilateral scientific cooperation no. 451-03-02141/2017-09/02, and by the Austrian Science Fund (FWF) through the SFB project “BeyondC” and a grant from the Foundational Questions Institute (FQXi) Fund. This publication was made possible through the support of the ID 61466 grant from the John Templeton Foundation, as part of the “The Quantum Information Structure of Spacetime (QISS)” Project (qiss.fr). The opinions expressed in this publication are those of the author(s) and do not necessarily reflect the views of the John Templeton Foundation.
\end{acknowledgments}


\appendix
\section{One Spin Particle}
\label{app:oneparticle}
\begin{figure}[htb]
	\centering
	\subfigure[\label{fig: OneDiracParticleLabFrame}
	Laboratory frame $\mathrm{C}$. ]{\includegraphics[scale=0.21]{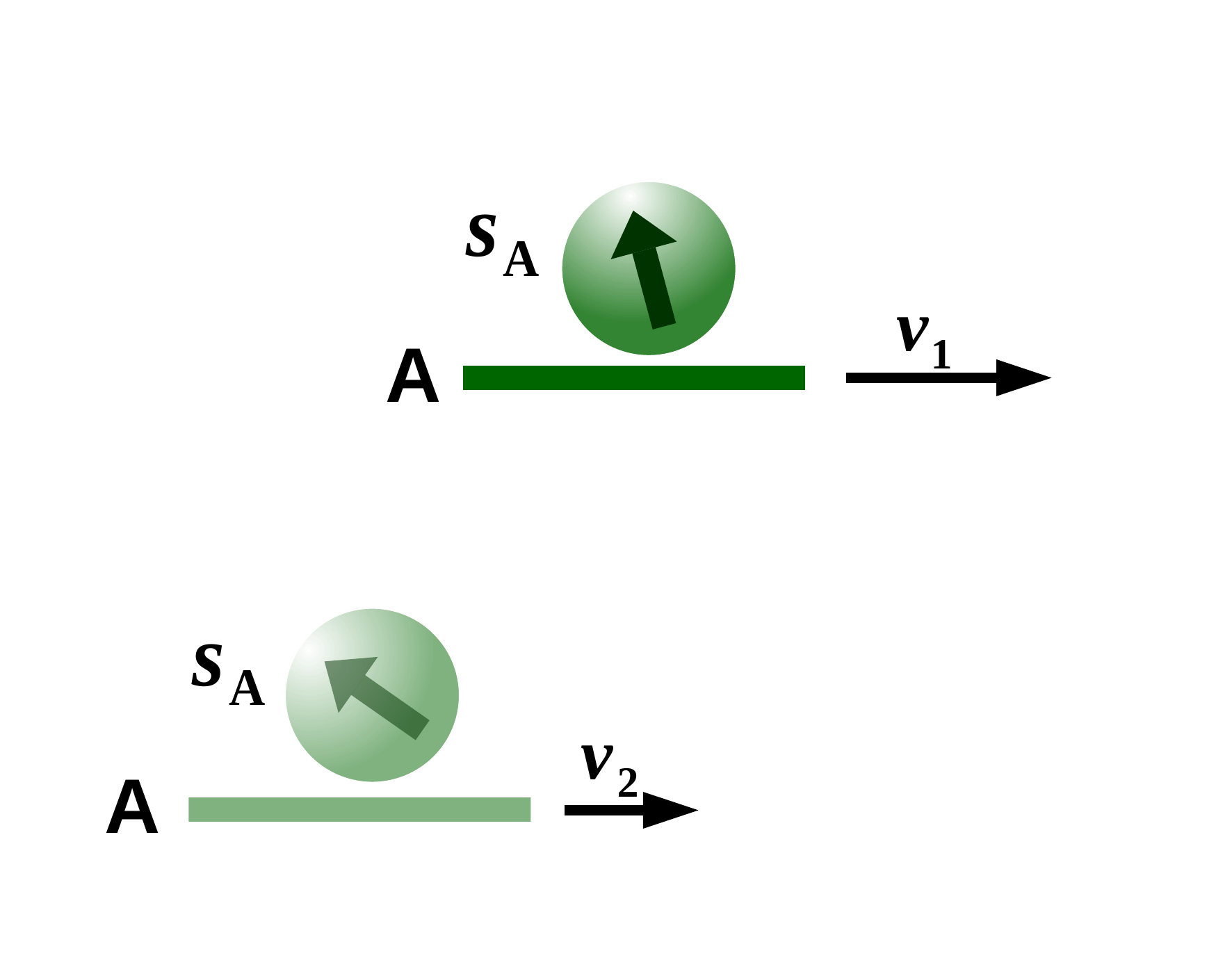}}
	\quad \quad \subfigure[\label{fig: OneDiracParticleRestFrame}
	Rest frame of the Dirac particle $\mathrm{A}$.]
	{\includegraphics[scale=0.21]{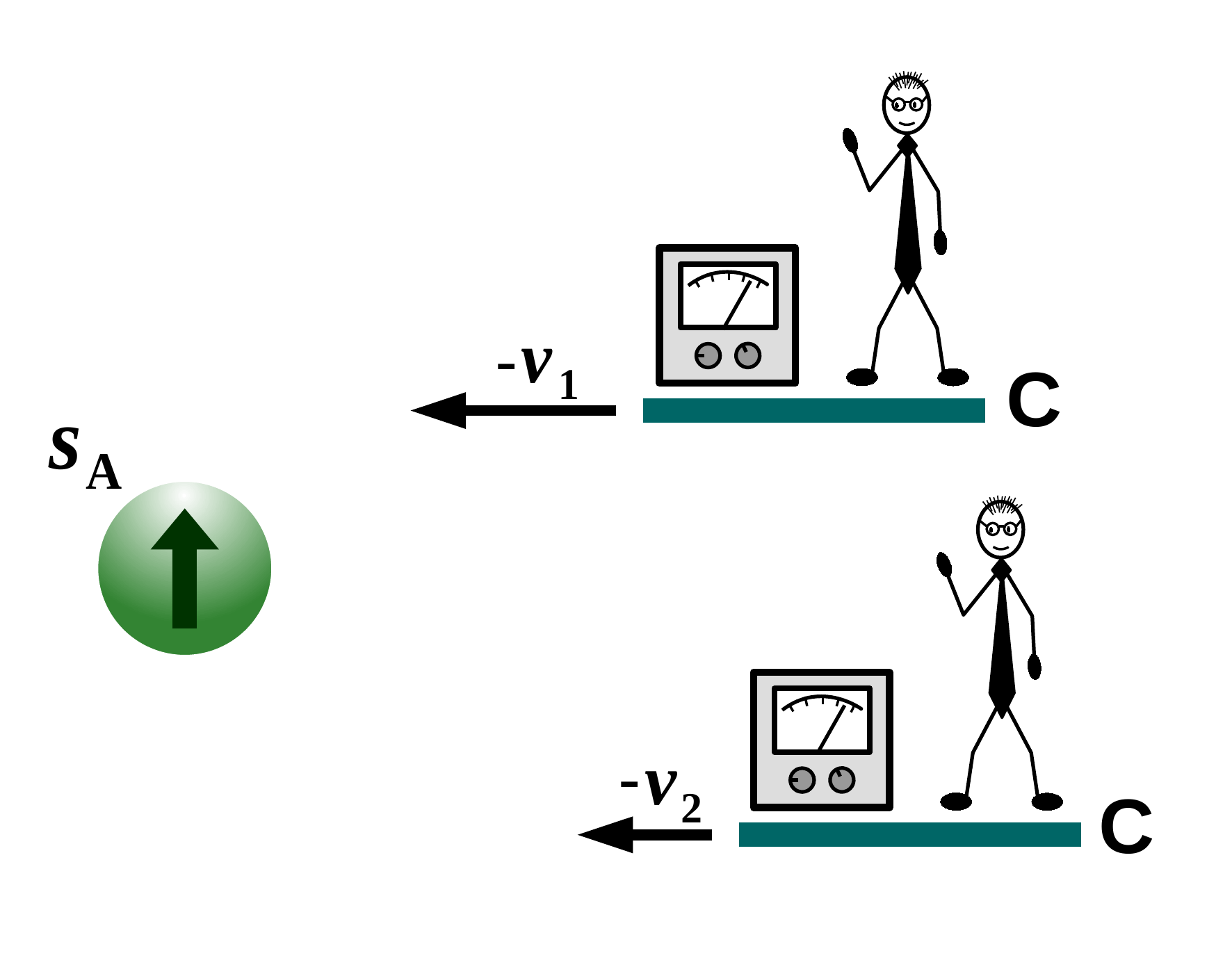}} 
	\caption{(a) In the laboratory perspective $\mathrm{C}$, the spin d.o.f.~$\spina$ is entangled with the momentum d.o.f.~$\mathrm{A}$ of the Dirac particle. (Pictorially, this is represented by the correlation between the transparency of the drawn symbols.) For simplicity, only two superposed velocities $v_1$ and $v_2$ are drawn. (b) Since the relative motion between all the systems is preserved, the laboratory is moving in a superposition of two sharp relativistic velocities $-v_1$ and $-v_2$ from the point of view of the particle's rest frame. In this QRF spin can be defined operationally by means of a Stern-Gerlach apparatus.}
	\label{fig: OneDiracParticle}
\end{figure}

\noindent
It is natural to define spin in the rest frame of a Dirac particle, where, in principle, spin can be tomographically verified by a series of standard Stern-Gerlach measurements. Through a QRF transformation (corresponding to a "superposition of Lorentz boosts") from the rest frame $\mathrm{A}$ to the laboratory frame $\mathrm{C}$ the relativistic spin operator $\hatvec{\Xi}$ is obtained. Consequently, the laboratory $\mathrm{C}$ describes the Dirac particle $\mathrm{A}$ with its external (momentum) and internal (spin) d.o.f.~which are labeled by $\mathrm{A}$ and $\spina$, respectively. This is graphically illustrated in Fig.~\ref{fig: OneDiracParticle}.

In mathematical terms, we start with a general separable state in the rest frame of the Dirac particle
\begin{equation}
\ket{\psi}^{|\mathrm{A}}_\mathrm{\spina \mathrm{C}} = \ket{\vec{\sigma}}_\spina \ket{\phi}_\mathrm{C} ,
\end{equation}
where $\ket{\vec{\sigma}}_\spina = \sum_a c_a \ket{a}_\spina$. Since the relative motion between the Dirac particle and the laboratory frame is superposed, the laboratory state from the perspective of the particle's rest frame is moving in a superposition of velocities which is described by the state
\begin{equation} \label{eq: laboratory state OneParticle}
\ket{\phi}_\mathrm{C} = \int \dmu_\mathrm{C}(\pi_\mathrm{C}) \phi(\pi_\mathrm{C}) \ket{\pi_\mathrm{C}}_\mathrm{C}
\end{equation}
with the (1+1)-momentum $\pi^\mu_\mathrm{C} = (\sqrt{\mc^2 c^2 + \pi_\mathrm{C}^2}, \pi_\mathrm{C})$ and the Lorentz-invariant integration measure $ \  \dmu_\mathrm{C}(\pi_\mathrm{C}) = \tfrac{\dx \pi_\mathrm{C} }{4 \pi \sqrt{m_\mathrm{C}^2 c^2 + \pi_\mathrm{C}^2}}$. For simplicity, the relative motion between $\mathrm{A}$ and $\mathrm{C}$ is restricted to be relativistic only in one dimension, while the spin of the Dirac particle is described in three spatial dimensions. 

The QRF transformation between the particle's and the laboratory's momentum d.o.f.~$\hat{S}_\mathrm{L}: \mathcal{H}^{|\mathrm{A}}_\spina \otimes \mathcal{H}^{|\mathrm{A}}_\mathrm{C} \mapsto \mathcal{H}^{|\mathrm{C}}_\mathrm{A} \otimes \mathcal{H}^{|\mathrm{C}}_\spina$ is given by
\begin{equation}
\hat{S}_\mathrm{L} = \mathcal{\hat{P}}^{(v)}_\mathrm{CA} \hat{U}_\spina (\hat{\pi}_\mathrm{C}) ,
\end{equation}
where $\mathcal{\hat{P}}^{(v)}_\mathrm{CA} = \mathcal{\hat{P}}_\mathrm{CA}  \expp{\tfrac{\im}{\hbar} \log\sqrt{\tfrac{m_\mathrm{A}}{m_\mathrm{C}}} \left(\hat{q}_\mathrm{C} \hat{\pi}_\mathrm{C} + \hat{\pi}_\mathrm{C} \hat{q}_\mathrm{C}\right)}$ is the generalized parity-swap operator mapping $\mathcal{\hat{P}}^{(v)}_\mathrm{CA} \hat{\pi}_\mathrm{C} \mathcal{\hat{P}}^{(v) \dagger}_\mathrm{CA} = - \tfrac{\mc}{\ma} \hat{p}_\mathrm{A}$ since $\mathcal{\hat{P}}_\mathrm{CA} \hat{\pi}_\mathrm{C} \mathcal{\hat{P}}_\mathrm{CA}^\dagger = -\hat{p}_\mathrm{A}$ and $\hat{U}_\spina(\hat{\pi}_\mathrm{C})$ is a unitary operator depending on the momentum of $\mathrm{C}$ and acting on the spin $\spina$ which transforms the spin from its rest frame to the laboratory frame. 

The action of the QRF transformation can be defined via the action on a basis of the total Hilbert space $\hat{S}_\mathrm{L} \ket{a}_\spina \ket{\pi}_\mathrm{C} = \ket{-\tfrac{\ma}{\mc} \pi; \Sigma(a)}_\aspina$. The quantum state in the laboratory frame is then
\begin{equation}
\begin{split}
\ket{\psi}^{|\mathrm{C}}_\aspina 
& = \hat{S}_\mathrm{L} \ket{\psi}^{|\mathrm{A}}_{\spina \mathrm{C}} \\
& = \sum_a c_a \int \dmu_\mathrm{A}(p) \ \phi\left(-\tfrac{\mc}{\ma} p\right) \ket{p; \Sigma(a)}_\aspina ,
\end{split}
\end{equation}
where $\ket{p; \Sigma(a)}_\aspina \equiv \hat{U}_\aspina(L_{p/\ma}) \ket{0; a}_\aspina$ and $\hat{U}_\aspina(L_{p/\ma})$ denotes a unitary representation of a pure Lorentz boost taking $k^\mu = (m_\mathrm{A} c, 0)$ to $p^\mu = (L_{p/\ma})^\mu_{\ \nu} k^\nu = (p^0, p) = (\sqrt{\ma^2 c^2 + p^2}, p)$. The pure Lorentz boost matrix is explicitly given in Appendix~\ref{app: Pure Lorentz Boosts}.

Most importantly, the relativistic spin operator in the laboratory frame
\begin{equation}
\hat{\Xi}^i_{\hat{p}_\mathrm{A}} = \hat{S}_\mathrm{L} \left(\hat{\sigma}_\spina^i \otimes \mathds{1}_\mathrm{C}\right)  \hat{S}_\mathrm{L}^\dagger
\end{equation}
satisfies the $\mathfrak{su}(2)$ algebra and reveals the same spin eigenvalue as in the particle's rest frame, i.e. $\hat{\Xi}^i_{\hat{p}_\mathrm{A}} \ket{p;\Sigma(a)}_\aspina = \sum_{a'} [\sigma^i]_{a'a} \ket{p;\Sigma(a')}_\aspina$ and $\sigma^i_\spina \ket{a}_\spina = \sum_{a'} [\sigma^i]_{a'a} \ket{a'}_\spina$, where $[\sigma^l]_{mn}$ denotes a matrix element of $\hat{\sigma}^l$. Moreover, notice that $\hat{\Xi}^i_{\hat{p}_\mathrm{A}} \ket{p;\Sigma(a)}_\aspina = \hat{\Xi}^i_{p} \ket{p;\Sigma(a)}_\aspina = \hat{U}_\aspina(L_{p}) (\mathds{1}_\mathrm{A} \otimes \hat{\sigma}^i_\spina) \hat{U}_\aspina^\dagger(L_{p}) \ket{p;\Sigma(a)}_\aspina$. It is worth mentioning that $\hatvec{\Xi}$ can be written in terms of the Pauli-Luba\'{n}ski spin operator, see Ref.~\cite{Giacomini2019Spin}.


\section{Pure Lorentz Boosts} \label{app: Pure Lorentz Boosts}
\noindent
The explicit matrix for pure Lorentz boosts, transforming to a reference frame which is moving with velocity $\vec{v}$ relative to the initial frame, is given by 
\begin{equation} \label{eq: Lorentz Matrix Velocity}
L_{\vec{v}} =
\begin{pmatrix}
\gamma	& \ \ \ \gamma \frac{\vec{v}^\top}{c} \\
\gamma \frac{\vec{v}}{c} & \mathds{1} + \frac{\gamma^2}{\gamma+1} \frac{\vec{v}\vec{v}^\top}{c^2}
\end{pmatrix}  
\end{equation}
where $\gamma \equiv (1-\frac{\vec{v}^2}{c^2})^{-1/2}$ \cite{Sexl2001a}. 
This matrix can be written completely in terms of the momentum $\vec{p}$, or more accurately the ratio $\vec{p}/m$, by substituting $p^\mu \equiv (p^0, \vec{p}) = m \gamma (c, \vec{v})$ in Eq.~(\ref{eq: Lorentz Matrix Velocity})
\begin{equation} \label{eq: Lorentz Matrix Momenta}
L_{\vec{p}} \equiv L_{\frac{\vec{p}}{m}} =
\begin{pmatrix}
\frac{p^0}{m c}	& \frac{\vec{p}^\top}{mc} \\
\frac{\vec{p}}{mc} & \mathds{1} + \frac{1}{\gamma + 1} \frac{\vec{p} \vec{p}^\top}{(m c)^2}
\end{pmatrix} 
\end{equation}
with $\gamma = \gamma_{\frac{\vec{p}}{m}} = \sqrt{1 + \frac{\vec{p}^2}{m^2 c^2}}$ and $p^0 = m c \gamma = \sqrt{m^2 c^2 + \vec{p}^2} = p^0(|\vec{p}|)$.

Notice that in Eq.~(\ref{eq: Lorentz Matrix Velocity}) and Eq.~(\ref{eq: Lorentz Matrix Momenta}) the transpose is explicitly denoted for compactness and correctness; however, not to overload the notation, the transpose is not denoted explicitly elsewhere.


\section{Correlation Tensor} \label{app: Correlation Tensor}
\noindent

In order to prove the violation of the CHSH-Bell inequality, we need to calculate the expectation value $E(\vec{x},\vec{y})$ of the joint spin measurement which is QRF-independent thanks to unitarity of the QRF transformation. Thus, we choose to calculate the expectation value in the rest frame of the Dirac particle $\mathrm{A}$.
 
From the point of view of particle $\mathrm{A}$, the total state is described by
\begin{equation} \label{eq: Collinear State}
\ket{\psi}^{|\mathrm{A}}_{\spina \bspinb C} = \ket{\eta}_{\spina \bspinb} \ket{\phi}_\mathrm{C}
\end{equation}
where 
\begin{align} 
	&\ket{\eta}_{\spina \bspinb} = \sum_{a,b} c_{ab} \int \dmu_\mathrm{B}(\pi_\mathrm{B}) \ \eta(\pi_\mathrm{B}) \ket{a}_\spina \ket{\pi_\mathrm{B}; \Sigma(b)}_\bspinb, \label{eq: Collinear State 1}\\
	&\ket{\phi}_\mathrm{C} = \int \dmu_\mathrm{C}(\pi_\mathrm{C}) \phi(\pi_\mathrm{C}) \ket{\pi_\mathrm{C}}_\mathrm{C}, \label{eq: Collinear State 2}
\end{align}
the (1+1)-momentum is $\pi^\mu_\mathrm{k} = (\sqrt{m_\mathrm{k}^2 c^2 + \pi_\mathrm{k}^2}, \pi_\mathrm{k})$ and the Lorentz-invariant integration measure $ \  \dmu_\mathrm{k}(\pi_\mathrm{k}) = \tfrac{\dx \pi_\mathrm{k} }{4 \pi \sqrt{m_\mathrm{k}^2 c^2 + \pi_\mathrm{k}^2}}$, with $\mathrm{k}= \mathrm{B, C}$.

The joint spin measurement in $\mathrm{A}$'s rest frame is described by
\begin{equation}
\vec{x} \cdot \hatvec{\sigma}_\spina \otimes \vec{y} \cdot \hatvec{\Xi}_{\hat{\pi}_\mathrm{B}} \otimes \mathds{1}_\mathrm{C} = \sum_{i,j} x^i y^j  \ \hat{\sigma}_{\spina}^i \otimes \hat{\Xi}_{\hat{\pi}_\mathrm{B}}^j \otimes \mathds{1}_\mathrm{C} , \label{eq: Collinear Observable}
\end{equation}
where $\vec{x} = (x^1, x^2, x^3)$ and $\vec{y} = (y^1, y^2, y^3)$ denote the measurement settings.

For the state and observable above, Eq.~(\ref{eq: Collinear State}) and (\ref{eq: Collinear Observable}), the expectation value is given by $E(\vec{x},\vec{y}) = \sum_{i,j} x^i y^j \ T^{ij}$, where
\begin{equation}
T^{ij} = \bra{\psi} \hat{\sigma}_{\spina}^i \otimes \hat{\Xi}_{\hat{\pi}_\mathrm{B}}^j \otimes \mathds{1}_\mathrm{C} \ket{\psi} = \bra{\eta} \hat{\sigma}_{\spina}^i \otimes \hat{\Xi}_{\hat{\pi}_\mathrm{B}}^j \ket{\eta} ,
\end{equation}
with $\ket{\psi} = \ket{\psi}^{|\mathrm{A}}_{\spina \bspinb C}$ and $\ket{\eta} = \ket{\eta}_{\spina \bspinb}$.

With the considered class of states, Eq.~(\ref{eq: Collinear State 1}), it follows that
\begin{equation}
\begin{split}
& \hat{\sigma}_{\spina}^i \otimes \hat{\Xi}_{\hat{\pi}_\mathrm{B}}^j \ket{\eta}_{\spina \bspinb} \\
& = \sum_{a,b} c_{ab} \! \int \dmu_\mathrm{B}(\pi_\mathrm{B}) \ \eta(\pi_\mathrm{B}) \  \hat{\sigma}_{\spina}^i \ket{a}_\spina \! \! \otimes  \hat{\Xi}_{\hat{\pi}_\mathrm{B}}^j \ket{\pi_\mathrm{B}; \Sigma(b)}_\bspinb ,
\end{split} 
\end{equation}
where $\hat{\sigma}_{\spina}^i \ket{a}_\spina = \sum_{a'} [\sigma^i]_{a'a} \ket{a'}_\spina$ and
\begin{equation}
\begin{split}
\hat{\Xi}_{\hat{\pi}_\mathrm{B}}^j 
& \ket{\pi_\mathrm{B}; \Sigma(b)}_\bspinb = \hat{\Xi}_{\pi_\mathrm{B}}^j \ket{\pi_\mathrm{B}; \Sigma(b)}_\bspinb \\
& = \hat{U}_\bspinb(L_{\pi_\mathrm{B}}) \left(\mathds{1}_\mathrm{B} \otimes \hat{\sigma}^j_\spinb\right) \hat{U}^\dagger_\bspinb(L_{\pi_\mathrm{B}}) \ket{\pi_\mathrm{B}; \Sigma(b)}_\bspinb \\
& = \sum_{b'} [\sigma^j]_{b'b} \ket{\pi_\mathrm{B}; \Sigma(b')}_\bspinb
\end{split}
\end{equation}
with $[\sigma^l]_{mn}$ denoting a matrix element of the Pauli operator $\hat{\sigma}^l$. By using the orthogonality relations  $\braket{\bar{a}|a'} = \delta_{\bar{a}a'}$ as well as $\braket{\bar{\pi}_\mathrm{B}; \Sigma(\bar{b})| \pi_\mathrm{B}; \Sigma(b')} = (2 \pi) 2 \bar{\pi}^0_\mathrm{B} \delta_{\bar{b}b'} \delta(\pi_\mathrm{B}-\bar{\pi}_\mathrm{B})$, the correlation tensor is given by
\begin{equation}
T^{ij} = \sum_{a,a',b,b'} c^*_{a'b'} c_{ab} [\sigma^i]_{a'a} [\sigma^j]_{b'b} \int \dmu_\mathrm{B}(\pi_\mathrm{B}) \ |\eta(\pi_\mathrm{B})|^2 .
\end{equation}
With the help of the normalization constraint
\begin{equation} \label{eq: normalization constraint}
\begin{split}
& \left[\braket{\eta|\eta} \right]^{|\mathrm{A}} = \sum_{a,b} |c_{ab}|^2 \ \int \dmu_\mathrm{B}(\pi_\mathrm{B}) \ |\eta(\pi_\mathrm{B})|^2 = 1 \\
& \Leftrightarrow  \sum_{a,b} |c_{ab}|^2=1 \quad \textmd{and} \quad \int \dmu_\mathrm{B}(\pi_\mathrm{B}) \ |\eta(\pi_\mathrm{B})|^2 = 1,
\end{split}
\end{equation}
we finally obtain
\begin{equation}
T^{ij} = \sum_{a,a',b,b'} c^*_{a'b'} c_{ab} [\sigma^i]_{a'a} [\sigma^j]_{b'b} .
\end{equation}
 
For the entangled state
\begin{equation} 
\begin{split}
& \ket{\eta^{-}}_{\spina \bspinb} \\
& \ \ = \sum_{\lambda=\pm z} c_\lambda \int \dmu_\mathrm{B}(\pi_\mathrm{B}) \ \eta(\pi_\mathrm{B}) \ket{\lambda}_\spina \ket{\pi_\mathrm{B}; \Sigma(-\lambda)}_\bspinb
\end{split}
\end{equation}
with $c_{\pm z} = \pm 1 / \sqrt{2}$ the correlation tensor reduces to
\begin{equation}
T^{ij} = \sum_{\lambda,\lambda'} c^*_{\lambda'} c_{\lambda} \ [\sigma^i]_{\lambda' \lambda} [\sigma^j]_{-\lambda',-\lambda} = - \delta^{ij} ,
\end{equation}
where $\lambda, \lambda' = \pm z$ and $[\sigma^l]_{mn}$ denotes a matrix element of $\hat{\sigma}^l$. Thus, $E(\vec{x},\vec{y}) = \sum_{i,j} x^i y^j \ T^{ij} = - \vec{x} \cdot \vec{y}$ as in the non-relativistic treatment, see Appendix~\ref{app: CHSH-Bell Inequality}.


\section{CHSH-Bell Inequality} \label{app: CHSH-Bell Inequality}

The Clauser-Horne-Shimony-Holt version of Bell's inequalities~\cite{CHSH1969}, referred to as \textit{CHSH-Bell inequality}, is considered. For joint measurements on two space-like separated systems, say $\spina$ and $\spinb$, with the outcomes $\pm 1$ for each measurement, the CHSH-Bell inequality sets a bound on the correlations of the two systems according to
\begin{equation} \label{eq: CHSH inequality}
|S| = | E(\vec{x}_1,\vec{y}_1) + E(\vec{x}_1,\vec{y}_2) + E(\vec{x}_2,\vec{y}_1) - E(\vec{x}_2,\vec{y}_2) | \leq 2,
\end{equation}
where $E(\vec{x}_i, \vec{y}_j)$ denotes the expectation value of the joint measurement on $\spina$ and $\spinb$, and the measurement settings $\vec{x}_1$ and $\vec{x}_2$ ($\vec{y}_1$ and $\vec{y}_2$) refer to $\spina$ ($\spinb$). In non-relativistic quantum mechanics, a joint spin measurement is described by the observable $\vec{x} \cdot \hatvec{\sigma}_\spina \otimes \vec{y} \cdot \hatvec{\sigma}_\spinb$, where $\hatvec{\sigma} \equiv (\hat{\sigma}^1,\hat{\sigma}^2,\hat{\sigma}^3)$ denotes the Pauli operator and $\vec{x} = (x^1, x^2, x^3)$ as well as $\vec{y}=(y^1,y^2,y^3)$ with $|\vec{x}| = |\vec{y}| = 1$ are Bloch vectors referring to the orientation of the Stern-Gerlach apparatuses. For the singlet state $\ket{\Psi^-}_{\spina \spinb} = (\ket{+z}_\spina \ket{-z}_\spinb - \ket{-z}_\spina \ket{+z}_\spinb)/\sqrt{2}$ the expectation value for a joint spin measurement is given by 
\begin{equation} \label{eq: Expectation value spin non rel}
\begin{split}
E(\vec{x},\vec{y}) 
& = \braket{\Psi^-|\vec{x} \cdot \hatvec{\sigma}_\spina \otimes \vec{y} \cdot \hatvec{\sigma}_\spinb | \Psi^-} = - \vec{x} \cdot \vec{y}
\end{split}
\end{equation}
and an optimal choice of measurement settings, e.g. $\vec{x}_1 = (0,1,1)/\sqrt{2}$, $\vec{x}_2 = (0,1,-1)/\sqrt{2}$, $\vec{y}_1 = (0,1,0)$ and $\vec{y}_2 = (0,0,1)$, leads to $|S| = 2\sqrt{2}$ which maximally violates the CHSH-Bell inequality.


\section{Transformation of the Bell Observables for Collinear Motion}
\label{app: BellObservableTrafoCollinear}
\noindent
In this section, we derive the general form of the Bell observable in the laboratory frame $\hat{O}^{ij}_{|\mathrm{C}} = \hat{S}_2 \hat{O}^{ij}_{|\mathrm{A}} \hat{S}_2^\dagger$, where $\hat{O}^{ij}_{|\mathrm{A}} \equiv \hat{\sigma}_{\spina}^i \otimes \hat{\Xi}_{\hat{\pi}_\mathrm{B}}^j \otimes \mathds{1}_\mathrm{C}$. For this purpose, the considered class of states in the rest frame of $\mathrm{A}$ is kept general as in Eq.~(\ref{eq: Collinear State}). Due to the unitarity of $\hat{S}_2$ we find
\begin{equation} \label{eq: Action Bell observable rest frame}
\begin{split}
& \hat{O}^{ij}_{|\mathrm{C}} \ket{\psi}^\mathrm{|C}_{\aspina \bspinb} \\
& \ \ = \hat{S}_2 \hat{O}^{ij}_{|\mathrm{A}} \hat{S}^\dagger_2 \hat{S}_2 \ket{\psi}^{|\mathrm{A}}_{\spina \bspinb \mathrm{C}} = \hat{S}_2 \hat{O}^{ij}_{|\mathrm{A}} \ket{\psi}^{|\mathrm{A}}_{\spina \bspinb \mathrm{C}} \\
& \ \ = \hat{S}_2 \left(\left[\hat{\sigma}_{\spina}^i \otimes \hat{\Xi}_{\hat{\pi}_\mathrm{B}}^j \ket{\eta}_{\spina \bspinb}\right] \otimes  \ket{\phi}_\mathrm{C}\right) ,
\end{split}
\end{equation}
where 
\begin{equation}
	\begin{split}
	\ket{\psi}^{|\mathrm{C}}_{\aspina \bspinb} = \sum_{a,b} & c_{ab} \int \dmu_\mathrm{A}(p_\mathrm{A}) \ \ \! \dmu_\mathrm{B}(p_\mathrm{B}) \ \eta(L^{-1} p_\mathrm{B}) \\ 
	& \phi\left(- \tfrac{\mc}{\ma} p_\mathrm{A}\right) \ket{p_\mathrm{A};\Sigma(a)}_\aspina \ket{p_\mathrm{B}; \Sigma(b)}_\bspinb .
	\end{split}
	\end{equation}

The action of the spin operators is given by $\sigma^i_\spina \ket{a}_\spina = \sum_{a'} [\sigma^i]_{a'a} \ket{a'}_\spina$ and $\hat{\Xi}^j_{\hat{\pi}_\mathrm{B}} \ket{\pi_\mathrm{B};\Sigma(b)}_\bspinb = \hat{\Xi}^j_{\pi_\mathrm{B}} \ket{\pi_\mathrm{B};\Sigma(b)}_\bspinb = \hat{U}_\bspinb(L_{\pi_\mathrm{B}}) (\mathds{1}_\mathrm{B} \otimes \hat{\sigma}^j_\spinb) \hat{U}_\bspinb^\dagger(L_{\pi_\mathrm{B}}) \ket{\pi_\mathrm{B};\Sigma(b')}_\bspinb = \sum_{b'} [\sigma^j]_{b'b} \ket{\pi_\mathrm{B};\Sigma(b')}_\bspinb$. Thus, we obtain
\begin{equation}
\begin{split}
\hat{O}&^{ij}_{|\mathrm{C}} \ket{\psi}^\mathrm{|C}_{\aspina \bspinb} \\
&= \sum_{a,a',b,b'} c_{ab} \ [\sigma^i]_{a'a} [\sigma^j]_{b'b} \int \dmu_\mathrm{A}(p_\mathrm{A}) \ \ \! \dmu_\mathrm{B}(p_\mathrm{B}) \\
& \ \ \eta(L^{-1} p_\mathrm{B}) \phi\left(- \tfrac{\mc}{\ma} p_\mathrm{A}\right) \ket{p_\mathrm{A};\Sigma(a')}_\aspina \ket{p_\mathrm{B}; \Sigma(b')}_\bspinb ,
\end{split}
\end{equation}
where we can now rewrite $\sum_{\lambda'} [\sigma^k]_{\lambda' \lambda} \ket{p;\Sigma(\lambda')} = \hat{\Xi}^k_{p} \ket{p;\Sigma(\lambda)} = \hat{\Xi}^k_{\hat{p}} \ket{p;\Sigma(\lambda)}$.

Overall, we find
\begin{equation}
\hat{O}^{ij}_{|\mathrm{C}} \ket{\psi}^\mathrm{|C}_{\aspina \bspinb} = \hat{\Xi}^i_{\hat{p}_\mathrm{A}} \otimes \hat{\Xi}^j_{\hat{p}_\mathrm{B}} \ket{\psi}^\mathrm{|C}_{\aspina \bspinb}
\end{equation}
thanks to the operator index of $\hat{\Xi}^l_{\hat{p}_\mathrm{k}}$, where $l = i,j$ and $\mathrm{k} =\mathrm{A}, \mathrm{B}$.

\section{Transformation of the Bell Observables for Non-collinear Motion}
\label{app: BellObservableTrafoNONCollinear}

In the following, we derive the form of the Bell observable for the non-collinear case as seen in the laboratory frame $\hat{G}^{\vec{x} \vec{y}}_{|\mathrm{C}} \equiv \hat{S}_2 (\vec{x} \cdot \hatvec{\sigma}_{\spina} \otimes \vec{y} \cdot \hatvec{\Xi}_{\hatvec{\pi}_\mathrm{B}} \otimes \mathds{1}_\mathrm{C}) \hat{S}_2^\dagger$ via its action on the laboratory state, i.e.~via $\hat{G}^{\vec{x} \vec{y}}_{|\mathrm{C}} \ket{\vec{\psi}}^{|\mathrm{C}}_{\aspina \bspinb}$. The calculation is similar to the one in Appendix~\ref{app: BellObservableTrafoCollinear}, but with an additional (Wigner) rotation of the spin d.o.f.~$\spinb$.
	
Thanks to the unitarity of $\hat{S}_2$, we get
\begin{equation}
	\begin{split}
	& \hat{G}^{\vec{x} \vec{y}}_{|\mathrm{C}} \ket{\vec{\psi}}^{|\mathrm{C}}_{\aspina \bspinb} \\
	& \quad = \hat{S}_2 \left(\vec{x} \cdot \hatvec{\sigma}_{\spina} \otimes \vec{y} \cdot \hatvec{\Xi}_{\hatvec{\pi}_\mathrm{B}} \otimes \mathds{1}_\mathrm{C}\right) \ket{\psi}^{|\mathrm{A}}_{\spina \bspinb \mathrm{C}} ,
	\end{split}
\end{equation}
where we consider the state in $\mathrm{A}$'s rest frame
\begin{equation}
\ket{\vec{\psi}}^{|\mathrm{A}}_{\spina \bspinb C} = \ket{\vec{\eta}}_{\spina \bspinb} \ket{\vec{\phi}}_\mathrm{C}
\end{equation}
with
\begin{equation}
\ket{\vec{\eta}}_{\spina \bspinb} = \sum_{a,b} c_{ab} \int \dmu_\mathrm{B}(\pi_\mathrm{B}) \ \eta(\vec{\pi}_\mathrm{B}) \ket{a}_\spina \ket{\vec{\pi}_\mathrm{B}; \Sigma(b)}_\bspinb
\end{equation}
and
\begin{equation}
\ket{\vec{\phi}}_\mathrm{C} = \int \dmu_\mathrm{C}(\pi_\mathrm{C}) \phi(\vec{\pi}_\mathrm{C}) \ket{\vec{\pi}_\mathrm{C}}_\mathrm{C}
\end{equation}
which leads to the corresponding state in the laboratory frame
\begin{equation}
\begin{split}
& \ket{\vec{\psi}}^{|\mathrm{C}}_{\aspina \bspinb} = \hat{S}_2 \ket{\vec{\psi}}^{|\mathrm{A}}_{\spina \bspinb \mathrm{C}} \\
& \ \ = \sum_{a,b} c_{ab} \int \dmu_\mathrm{A}(p_\mathrm{A}) \ \ \! \dmu_\mathrm{B}(p_\mathrm{B}) \ \eta(L^{-1} \vec{p}_\mathrm{B}) \\
& \quad \quad \quad \quad \phi\left(- \tfrac{\mc}{\ma} \vec{p}_\mathrm{A}\right)  \ket{\vec{p}_\mathrm{A};\Sigma(a)}_\aspina \ket{\vec{p}_\mathrm{B}; \Sigma(R_{\vec{\Omega}}(b))}_\bspinb .
\end{split}
\end{equation}

Thus, we obtain
\begin{equation}
\begin{split}
& \hat{G}^{\vec{x} \vec{y}}_{|\mathrm{C}} \ket{\vec{\psi}}^{|\mathrm{C}}_{\aspina \bspinb} \\
& \ \ = \hat{S}_2 \sum_{i,j,a,b,a',b'} x_i y_j \ c_{ab} \ [\sigma^i]_{a'a} [\sigma^j]_{b'b} \\
& \ \ \hspace{1.3cm} \int \dmu_\mathrm{B}(\pi_\mathrm{B}) \ \ \! \dmu_\mathrm{C}(\pi_\mathrm{C}) \ 
\eta(\vec{\pi}_\mathrm{B}) \phi\left(\vec{\pi}_\mathrm{C}\right) \\
& \ \ \hspace{3.2cm} \ket{a'}_\spina \ket{\vec{\pi}_\mathrm{B}; \Sigma(b')}_\bspinb \ket{\vec{\pi}_\mathrm{C}}_\mathrm{C} \\
& \ \ = \sum_{i,j,a,b,a',b'} x_i y_j \ c_{ab} \ [\sigma^i]_{a'a} [\sigma^j]_{b'b} \\
& \ \ \hspace{0.5cm} \int \dmu_\mathrm{A}(p_\mathrm{A}) \ \ \! \dmu_\mathrm{B}(p_\mathrm{B}) \ \eta(L^{-1} \vec{p}_\mathrm{B}) \phi\left(- \tfrac{\mc}{\ma} \vec{p}_\mathrm{A}\right) \\ 
& \ \ \hspace{2.2cm} \ket{\vec{p}_\mathrm{A};\Sigma(a')}_\aspina \ket{\vec{p}_\mathrm{B}; \Sigma(R_{\vec{\Omega}}(b'))}_\bspinb \\
\end{split}
\end{equation}
where for the Dirac particle $\mathrm{A}$ we know already that $\sum_{a'} [\sigma^i]_{a'a} \ket{\vec{p}_\mathrm{A};\Sigma(a')}_\aspina = \hat{\Xi}^i_{\hatvec{p}_\mathrm{A}} \ket{\vec{p}_\mathrm{A};\Sigma(a)}_\aspina$. For Dirac particle $\mathrm{B}$, we find
\begin{equation}
\begin{split}
\sum_{j,b'} & y_j [\sigma^j]_{b'b} \ket{\vec{p}_\mathrm{B}; \Sigma(R_{\vec{\Omega}}(b'))}_\bspinb \\
& = \sum_{j,b'} y_j [\sigma^j]_{b'b} \hat{U}(L_{\vec{p}_\mathrm{B}})\ket{0; R_{\vec{\Omega}}(b')}_\bspinb \\
& = \sum_{j,b'} y_j [\sigma^j]_{b'b} \hat{U}(L_{\vec{p}_\mathrm{B}}) \left[\mathds{1}_\mathrm{B} \otimes \hat{R}_\spinb(\vec{\Omega})\right] \ket{0;b'}_\bspinb \\
& = \hat{U}(L_{\vec{p}_\mathrm{B}}) \left[\mathds{1}_\mathrm{B} \otimes \hat{R}_\spinb(\vec{\Omega})\right] (\mathds{1}_\mathrm{B} \otimes \vec{y} \cdot \hatvec{\sigma}_\spinb) \ket{0;b}_\bspinb
\end{split}
\end{equation}
and by inserting the unit element $\mathds{1}_{\bspinb} = [\mathds{1}_\mathrm{B} \otimes \hat{R}^\dagger_\spinb(\vec{\Omega})] \hat{U}^\dagger(L_{\vec{p}_\mathrm{B}}) \hat{U}(L_{\vec{p}_\mathrm{B}}) [\mathds{1}_\mathrm{B} \otimes \hat{R}_\spinb(\vec{\Omega})]$ in front of the state we get
\begin{equation}
\begin{split}
\sum_{j,b'} & y_j [\sigma^j]_{b'b} \ket{\vec{p}_\mathrm{B}; \Sigma(R_{\vec{\Omega}}(b'))}_\bspinb \\
& = \hat{U}(L_{\vec{p}_\mathrm{B}}) \left[\mathds{1}_\mathrm{B} \otimes \hat{R}_\spinb(\vec{\Omega}) (\vec{y} \cdot \hatvec{\sigma}_\spinb) \hat{R}^\dagger_\spinb(\vec{\Omega})\right] \hat{U}^\dagger(L_{\vec{p}_\mathrm{B}}) \\
& \hspace{5.1cm} \ket{\vec{p}_\mathrm{B}; \Sigma(R_{\vec{\Omega}}(b))}_\bspinb .
\end{split}
\end{equation}
It is straightforward to show that $\hat{R}_\spinb(\vec{\Omega}) (\vec{y} \cdot \hatvec{\sigma}_\spinb) \hat{R}^\dagger_\spinb(\vec{\Omega}) = \vec{y}^{R} \cdot \hatvec{\sigma}_\spinb$ where
\begin{equation} \label{eq: Rotated Setting}
\begin{split}
\vec{y}^{R} & = \vec{y}^{R}(\vec{\Omega}) \\
& = \vec{y} \cos\Omega + \vec{n} (\vec{n}\cdot\vec{y})(1-\cos\Omega) + (\vec{n} \times \vec{y}) \sin\Omega
\end{split}
\end{equation} 
and $\vec{\Omega} = \Omega \vec{n}$. This means that the measurement setting for the Dirac particle $\mathrm{B}$ in the laboratory frame is rotated with respect to the setting in the rest frame of $\mathrm{A}$. With this, it follows that
\begin{equation}
\begin{split}
& \sum_{j,b'} y_j [\sigma^j]_{b'b} \ket{\vec{p}_\mathrm{B}; \Sigma(R_{\vec{\Omega}}(b'))}_\bspinb \\
& \ \ = \hat{U}(L_{\vec{p}_\mathrm{B}}) (\mathds{1}_\mathrm{B} \otimes \vec{y}^R \cdot \hatvec{\sigma}_\spinb) \hat{U}^\dagger(L_{\vec{p}_\mathrm{B}}) \ket{\vec{p}_\mathrm{B}; \Sigma(R_{\vec{\Omega}}(b))}_\bspinb \\
& \ \ = \sum_j y_j^R \ \hat{U}(L_{\vec{p}_\mathrm{B}}) (\mathds{1}_\mathrm{B} \otimes \hat{\sigma}^j_\spinb) \hat{U}^\dagger(L_{\vec{p}_\mathrm{B}}) \ket{\vec{p}_\mathrm{B}; \Sigma(R_{\vec{\Omega}}(b))}_\bspinb \\
& \ \ = \sum_j y_j^R \ \hat{\Xi}^j_{\hatvec{p}_\mathrm{B}} \ket{\vec{p}_\mathrm{B}; \Sigma(R_{\vec{\Omega}}(b))}_\bspinb \\
& \ \ = \vec{y}^R \cdot \hatvec{\Xi}_{\hatvec{p}_\mathrm{B}} \ket{\vec{p}_\mathrm{B}; \Sigma(R_{\vec{\Omega}}(b))}_\bspinb
\end{split}
\end{equation}
and with $\vec{y}^R = \vec{y}^R(\vec{\Omega}) = \vec{y}^R(\vec{p}_\mathrm{A},\vec{p}_\mathrm{B})$ we obtain
\begin{equation} \label{eq: Wigner Rot Observable and State}
\begin{split}
& \hat{G}^{\vec{x} \vec{y}}_{|\mathrm{C}} \ket{\vec{\psi}}^{|\mathrm{C}}_{{\aspina \bspinb}} \\
& = \sum_{a,b} c_{ab} \int \dmu_\mathrm{A}(p_\mathrm{A}) \ \ \! \dmu_\mathrm{B}(p_\mathrm{B}) \ \eta(L^{-1} \vec{p}_\mathrm{B}) \phi\left(- \tfrac{\mc}{\ma} \vec{p}_\mathrm{A}\right) \\
& \quad \quad \quad \quad \quad \quad \vec{x} \cdot \hatvec{\Xi}_{\hatvec{p}_\mathrm{A}} \ket{\vec{p}_\mathrm{A};\Sigma(a)}_\aspina \otimes \\
& \quad \quad \quad \quad \quad \quad \quad \quad \vec{y}^R(\vec{p}_\mathrm{A},\hatvec{p}_\mathrm{B}) \cdot \hatvec{\Xi}_{\hatvec{p}_\mathrm{B}} \ket{\vec{p}_\mathrm{B}; \Sigma(R_{\vec{\Omega}}(b))}_\bspinb,
\end{split}
\end{equation}
where $\vec{p}_\mathrm{B}$ has been promoted to an operator which is possible since $\hatvec{\Xi}_{\hatvec{p}_\mathrm{B}}$ does not change the momentum of the Dirac particle $\mathrm{B}$. However, if we promote $\vec{p}_\mathrm{A} \to \hatvec{p}_\mathrm{A}$ in the argument of $\vec{y}^R$, then we cannot split the Bell observable in the laboratory frame into two observables $\hat{G}^{\vec{x}}_{\mathrm{A} \spina} \in \mathcal{H}_\aspina$ and $\hat{G}^{\vec{y}^R}_{\mathrm{B}\spinb} \in \mathcal{H}_\bspinb$ acting on particle $\mathrm{A}$ and $\mathrm{B}$ separately, i.e.
\begin{equation}
\begin{split}
\hat{G}^{\vec{x} \vec{y}}_{|\mathrm{C}}
& \equiv \hat{S}_2 \left(\vec{x} \cdot \hatvec{\sigma}_{\spina} \otimes \vec{y} \cdot \hatvec{\Xi}_{\hatvec{\pi}_\mathrm{B}} \otimes \mathds{1}_\mathrm{C}\right) \hat{S}_2^\dagger \\
& = \sum_j \left( \vec{x} \cdot \hatvec{\Xi}_{\hatvec{p}_\mathrm{A}} \otimes \mathds{1}_\aspina \right) \\
& \hspace{1cm} \left(y_j^R(\hatvec{p}_\mathrm{A},\hatvec{p}_\mathrm{B}) \otimes \mathds{1}_\spina \otimes \mathds{1}_\spinb \right) \left(\mathds{1}_\aspina \otimes \hat{\Xi}^j_{\hatvec{p}_\mathrm{B}} \right) \\
& \neq \hat{G}^{\vec{x}}_{\aspina} \otimes \hat{G}^{\vec{y}^R}_{\bspinb}.
\end{split}
\end{equation}
where $y_j^R(\hatvec{p}_\mathrm{A},\hatvec{p}_\mathrm{B})$ refers to the $j$-th component of $\vec{y}^R(\hatvec{p}_\mathrm{A},\hatvec{p}_\mathrm{B}) = \vec{y} \cos[\Omega(\hatvec{p}_\mathrm{A},\hatvec{p}_\mathrm{B})] + \vec{n} (\vec{n} \cdot \vec{y}) (1-\cos[\Omega(\hatvec{p}_\mathrm{A},\hatvec{p}_\mathrm{B})]) + (\vec{n} \times \vec{y}) \sin[\Omega(\hatvec{p}_\mathrm{A},\hatvec{p}_\mathrm{B})]$.

\vspace{10cm}
\bibliography{RelQRF.bib}

\end{document}